\documentclass[intlimits,twoside,a4paper]{article} 
\usepackage{cite}
\usepackage{amsmath,amssymb}
\usepackage{graphicx,psfrag} 
\usepackage[T2A]{fontenc} 
\usepackage[cp1251]{inputenc} 
\usepackage{soul}
\usepackage{xcolor}
\usepackage[eqsecnum]{cmpj3}

\def\be{\begin{equation}}
\def\ee{\end{equation}}
\def\bea{\begin{eqnarray}}
\def\eea{\end{eqnarray}}

\def\CK{{\mathcal K}}

\def\CH{{\mathcal H}}

\def\CR{{\mathcal R}}

\def\kBT{{k_{\mathrm B}T}}
\def\msc{{m_{sc}}}

\def\vf{{\varphi}}
\newcommand{\Or}{\mathord{\mathrm{O}}} 
\newcommand{\eref}[1]{(\ref{#1})}
\newcommand{\Tr}{\mathop{\mathrm{Tr}}\nolimits}

\issue{2023}{26}{1}{13101}
\doinumber{10.5488/CMP.26.13101}

\title[Scaling behaviour under the influence of 
a homogeneous size-dependent perturbation]%
{Scaling behaviour under the influence of 
a homogeneous size-dependent perturbation
}
\author[L. Turban]{\framebox{L. Turban}\orcid{0000-0003-1051-4497}}

\address{ Universit\'e de Lorraine, CNRS, LPCT, F-54000 Nancy, France %; \email{loic.turban@univ-lorraine.fr}
}
  

\date{Received August 9, 2022, in final form October 7, 2022}

\begin{document}

\maketitle
\begin{abstract}
We study the finite-size scaling behaviour at the critical point, resulting from the addition of a homogeneous size-dependent perturbation, decaying as an inverse power of the system size. The scaling theory is first formulated in a general framework and then illustrated using three concrete problems for which exact results are obtained. 
 
\printkeywords
%
%\pacs 
\end{abstract}

\begin{verse}\it
\hspace{6.8cm}Count your age by friends, not years.\\
\hspace{7cm}Count your life by smiles, not tears.\\
\hspace{9cm} Birthday card, 1927.
\end{verse}

\section*{Foreword}
I first met Bertrand in the mid-eighties when he was still undergraduate. Around 87--88 he joined our small statistical physics group to prepare his PhD. It was a pleasant time when research did not mean competition but collaboration. All along the years Bertrand was very efficient in keeping this state of mind in our group. He was one of the main contributors to the development of external collaborations as this volume bears witness. 
It is a pleasure to take part in this nice initiative of our Ukrainian friends and to wish you, Bertrand, all the best for your 60th birthday.

\section{Introduction} 
At a second-order phase transition a perturbation is relevant (irrelevant) when its amplitude increases (decreases) under rescaling.
In the relevant case, the fixed point initially governing the critical behaviour of the system is unstable and a new behaviour sets in.
When the perturbation is irrelevant, the fixed point is stable and the critical behaviour remains unaffected. In between, when the perturbation amplitude does not evolve at all under rescaling, the perturbation is said to be truly marginal. Then, the critical behaviour of the perturbed system is generally governed by a line of fixed points in the critical surface, with varying critical exponents, parameterized by the value of perturbation amplitude.  

In the present work we study the critical behaviour associated with homogeneous size-dependent perturbations, decaying as some inverse power of the system size, which deviate from the general trend mentioned above in the marginal case.

This type of perturbation can be generated by a conformal transformation
in two dimensions (2d). Then, a conformally invariant infinite (semi-infinite)  critical system transforms into a strip with periodic (free) boundary conditions under the logarithmic transformation. As shown by 
Cardy~\cite{cardy84}, the transformation of the two-point correlation functions provides an explanation for the gap-exponent relation~\cite{pichard81,luck81,luck82,derrida82,nightingale83,privman84}. When the system is initially perturbed by a marginal radial defect, it transforms into a strip with a homogeneous, size-dependent 
perturbation~\cite{bariev91,bariev91a,turban91} (see appendix~\ref{conf} for details). 

Another example of the occurrence of a size-dependent contribution to the Hamiltonian, but of a quite different nature, is offered by (mean-field) fully-interacting systems with $N$ spins. The number of interacting pairs growing as $N^2$, the double sum over the sites has to be divided by $N$ in order to obtain a finite energy per site in the thermodynamic 
limit~\cite{baker63,kittel65,niemeijer70,vertogen72,sherrington75}.
Here, we shall  consider the change in the finite-size scaling behaviour when a size-dependent perturbation is added to the interaction amplitude $K$.

The outline of the paper is as follows: section~\ref{sec:2}, which contains the main results, presents the finite-size scaling behaviour resulting from the introduction of a homogeneous size-dependent perturbation in the three different cases of irrelevant, marginal, and relevant perturbations. This is illustrated with three exactly solvable examples: percolation in 1d (section~\ref{sec:3}), the Ising chain in a transverse field corresponding to a 2d classical system (section~\ref{sec:4}) and the Ising model on the fully-connected lattice (section~\ref{sec:5}). The results are discussed in section~\ref{sec:6}. The logarithmic transformation of a radial perturbation is presented in appendix~\ref{conf}. In appendix~\ref{App2}, we consider a perturbation of the 1d percolation problem in which the size of the system is replaced by the distance to the surface (Hilhorst--van Leeuwen perturbation). The two last appendices give some calculational details.

\section{Finite-size scaling}
\label{sec:2}
 
Let us consider a perturbed $d$-dimensional classical system with Hamiltonian $\CH$ such that:
\be
-\beta\CH=-\beta\CH_0+\Delta\sum_i\zeta_i\,,\quad \beta=\frac{1}{\kBT}\,.
\label{H}
\ee
The system is finite, with size $L\gg1$, in at least one of its $d$ dimensions. The scaling field associated with  the perturbation, $\Delta$, is size-dependent and given by
\be 
\Delta=\frac{A}{L^{\omega}}\,,\quad \omega>0\,,\quad A=\Or(1)\,.
\label{Delta}
\ee
$\Delta$ is conjugate to the local operator $\zeta$
with scaling dimension $x_\zeta$ and the perturbation acts on a subspace with dimension $d_\Delta$. Thus, $d_\Delta=1$ for a line defect and $d_\Delta=d$ when the perturbation extends over the whole system.

Under a change of the length scale by a factor $b>1$, such that $L'=L/b$, the perturbation  transforms according to:
\be
\Delta'=\frac{A'}{{L'}^{\omega}}=b^{y_\Delta}\Delta\,,\quad y_\Delta=d_\Delta-x_\zeta.
\label{Dy}
\ee
Taking into account the size-dependence of the perturbation~\eref{Delta}, the following behaviour is obtained for the amplitude\footnote{Hereafter we assume that $y_\Delta>0$.}
\be
A'=b^{y_\Delta-\omega}A.
\label{Ay}
\ee

Suppose now that $\CH_0$ is at the bulk 
critical point and let us look for the finite-size
scaling behaviour of a local operator $\vf$, with scaling dimension $x_\vf$, under the influence of $\Delta$ acting on $\zeta$. At bulk criticality~$\vf_c$ depends only on  
two variables, the size-dependent perturbation amplitude $\Delta$ and the system size $L$. It transforms according to
\be
{\vf'}_c=\vf_c(\Delta',L')=b^{x_\vf}\vf_c(\Delta,L),
\label{vf'}
\ee
so that:
\be  
\vf_c(\Delta,L)=b^{-x_\vf}\vf_c(b^{y_\Delta}\Delta,L/b).
\label{vf}
\ee
With $b=L$ one obtains the following finite-size behaviour
\be 
\vf_c(\Delta,L)=\Phi_c(A,L)=L^{-x_\vf}\vf_c(L^{y_\Delta-\omega}A,1)=L^{-x_\vf}\phi_\vf(u),
\quad u=L^{y_\Delta-\omega}A,
\label{phi}
\ee
where the scaling function $\phi_\vf(u)$ is a universal function of its argument~\cite{privman84}, with $\phi_\vf(0)$ giving the universal finite-size scaling amplitude of the unperturbed system:
\be
\Phi_c(0,L)=\phi_\vf(0)L^{-x_\vf}.
\label{phiun}
\ee
Depending on the value of $\omega$, three cases have to be considered:
\begin{itemize}
\item Irrelevant perturbation, $\omega>y_\Delta$: 
According to~\eref{Ay}, the perturbation amplitude decreases under rescaling, and the scaling function $\phi_\vf(u)$ in~\eref{phi} can be expanded in powers of $u$ giving:
\be 
\Phi_c(A,L)=L^{-x_\vf}\left[\phi_\vf(0)+\frac{A}{L^{\omega-y_\Delta}}\phi'(0)+\ldots\right].
\label{phiirr}
\ee
The leading contribution is the unperturbed one in~\eref{phiun} and the perturbation affects only the sub-leading corrections to scaling.

\item Marginal perturbation, $\omega=y_\Delta$:
Then, the perturbation amplitude $A$ is invariant under rescaling and the argument of the scaling function in~\eref{phi} no longer depends on $L$ so that:
\be 
\Phi_c(A,L)=\phi_\vf(A)L^{-x_\vf}.
\label{phimar}
\ee 
The scaling behaviour, $L^{-x_\vf}$, is the same as for the unperturbed system but the finite-size scaling amplitude, $\phi_\vf(A)$, is now continuously varying with $A$. It is actually a universal function of $A$. 

\item Relevant perturbation, $\omega<y_\Delta$:
In this case, the perturbation amplitude grows under rescaling. Let us consider the situation where $\omega=0$ which, according to~\eref{Delta} corresponds to a constant deviation~$A$ from the critical point. Then, 
\be
\Phi_c(A,L)=L^{-x_\vf}\phi_\vf(L^{y_\Delta}A),
\quad \omega=0.
\label{phirel-1}
\ee
In the thermodynamic limit, either $\lim_{L\to\infty}
\Phi_c(A,L)=0$ or the exponent of $L$ on the 
right-hand side of~\eref{phirel-1} vanishes. This occurs when 
\be
\phi_\vf(u)\sim |u|^\kappa,\quad |u|\gg1,\quad
\kappa=\frac{x_\vf}{y_\Delta}.
\label{phirel-2}
\ee
so that $\Phi_c(A,\infty)\sim|A|^{x_\vf/y_\Delta}$. 
When $0<\omega<y_\Delta$, using~\eref{phirel-2} in~\eref{phi}, 
one obtains:
\be 
\Phi_c(A,L)\sim L^{-\omega x_\vf/y_\Delta}|A|^{x_\vf/y_\Delta}.
\label{phirel-3}
\ee
\end{itemize}

\section{Bond percolation in 1d}
\label{sec:3}

To illustrate the results of the preceding section, let us start with a simple example, namely the bond percolation problem~\cite{shante71} on a 1d lattice with a size-dependent bond occupation probability.

\subsection{Percolation probability} 
We consider a finite chain with $L$ bonds between 
neighbouring sites. The bonds are independently occupied 
with probability:
\be  
\pi=p+\frac{A}{L^\omega}\,,\quad -pL^\omega \leqslant A\leqslant (1-p)L^\omega
\,,\quad \omega \geqslant 0.
\label{pi}
\ee
The order parameter for the percolation transition is the percolation probability. It is given by the probability $P_L(\pi)$ to find an open path from 0 to $L$. Thus, according to~\eref{pi}, we have:
\be 
P_L(\pi)=\left(p+\frac{A}{L^\omega}\right)^L.
\label{PpAL}
\ee

\subsection{Unperturbed system}
A change of the length scale by $b$ can be exactly realized {\it via} decimation~\cite{reynolds77}. In the process, $b$ successive bonds with occupation probability $p$ are replaced by a single bond with renormalized probability
\be 
p'=\CR_b(p)=p^b.
\label{p'}
\ee
This exact renormalization group transformation has two fixed points at $p^\star=\CR_b(p^\star)=$ 0 or 1. The unstable point corresponds to
the percolation threshold, $p_c=1$. A linearization of the transformation about the fixed point at $p^\star=1$ gives the transformation of the deviation from the percolation threshold:
\be 
p'-p_c=\left.\frac{\rd\CR_b(p)}{\rd p}\right|_{p=p_c=1}(p-p_c)
=b(p-p_c).
\label{scalp}
\ee
Thus, the dimension of the scaling field, $p-p_c$, is
\be
y_p=1.
\label{yp}
\ee
The percolation probability, $P_L(p)=\CR_L(p)=p^L$, jumps from 0 to 1 at $p_c$  when $L\to\infty$:
\be
 P_\infty(p)=
 \left\{
 \begin{array}{ll}
 0,&\quad p<1,\\
  1,&\quad p=p_c=1.
 \end{array}
 \right.
 \label{Pinf}
 \ee
The percolation transition is discontinuous in 1d. 

On a finite chain, the critical percolation probability is independent of the size, $P_L(p_c)=1$.
Then, according to~\eref{phiun}, the scaling dimension of the order parameter is
\be 
x_P=0,
\label{xP}
\ee
as expected at a discontinuity fixed point. The finite-size percolation probability reduces to its scaling function, $\phi_P(0)=1$. Given~\eref{scalp}, the scaling variable can be defined as
\be 
u=L(p-p_c)=L(p-1)\leqslant0,\quad p=1+\frac{u}{L},
\label{uperco}
\ee
leading to
\be 
P_L(p)=\left(1+\frac{u}{L}\right)^L=\exp\left(u-\frac{u^2}{2L}+\ldots\right)
=\re^{u}\left(1-\frac{u^2}{2L}+\ldots\right),
\label{PLu}
\ee
when $u^2/L\ll1$. The scaling function of the percolation probability 
\be 
\phi_P(u)= \re^{u},\quad u\leqslant0,\quad u^2/L\ll1,
\label{phiPu}
\ee
follows from this expression.

Note that the correlation function $\Gamma_n(p)$, which is the probability
that two sites at a distance $n$ belong to the same cluster, is simply given by
\be
\Gamma_n(p)=P_n(p)\approx \re^{-n(p_c-p)},\quad p_c-p\ll1,
\label{Gamma}
\ee
according to~\eref{uperco} and~\eref{PLu}. Thus, the correlation length, $\xi_p=(p_c-p)^{-1}$, diverges at $p_c$
with an exponent $\nu_p=1/y_p=1$, in agreement with~\eref{yp}.

%%%%%%%%%% FIG 1  %%%%%%%%%%%%%%%%%%%%%%%%%%%%%%%
\begin{figure}[]
\begin{center}
\includegraphics[width=8cm,angle=0]{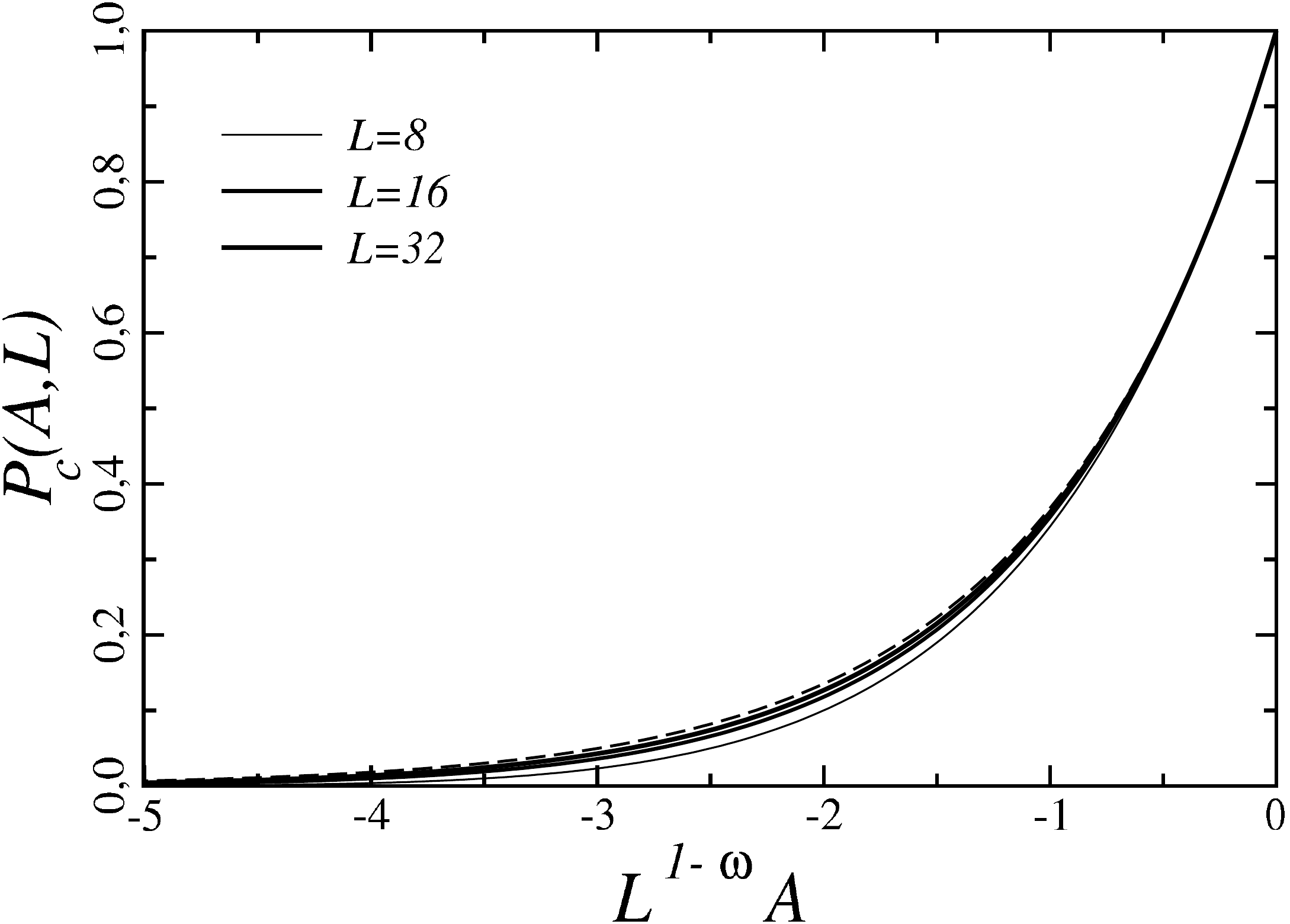}
\end{center}
\vglue -.5cm
\caption{Evolution of the critical percolation probability, given by~\eref{PcAL}, towards the universal scaling 
function $\phi_{P}(u)=\re^u$ (dashed line), for increasing values of the size $L$. There is no rescaling of the percolation probability since $x_P=0$ and the scaling variable is $u=L^{1-\omega}A$. }
\label{fig-1}
\end{figure}
%%%%%%%%%%%%%%%%%%%%%%%%%%%%%%%%%%%%%%%%%
 
\subsection{Perturbed system}

We now consider the perturbed 1d percolation problem with a bond 
occupation probability $\pi_c$ given by~\eref{pi} at $p=p_c=1$.
The transformation of the perturbation amplitude follows from~\eref{Ay} and~\eref{yp}:
\be
A'=b^{y_p-\omega}A=b^{1-\omega}A.
\label{scalA}
\ee
Thus, the perturbation is irrelevant when $\omega>1$,
marginal at $\omega=\omega^\star=1$ and relevant when 
$\omega<1$. 
The finite-size scaling variable is given by 
\be 
u=L(\pi_c-p_c)=L^{1-\omega}A\leqslant0,
\label{uAperco}
\ee
in agreement with~\eref{scalA}. The critical percolation probability
\be  
P_c(A,L)=P_L(\pi_c)=\left(1+\frac{u}{L}\right)^L,
\label{PcAL}
\ee
follows from~\eref{PLu}. Let us examine its scaling behaviour depending on the value of $\omega$.

\begin{itemize}

\item Irrelevant perturbation, $\omega>1$:

 Then, $|u|\ll1$ and a first-order expansion in~\eref{PLu} yields:
 \be 
 P_c(A,L)=1+\frac{A}{L^{\omega-1}}+\Or \left(L^{-2(\omega-1)}\right),
 \quad A<0.
 \label{Pcirr}
 \ee
 It should be compared to~\eref{phiirr} with $x_\vf=x_P=0$ 
 and $y_\Delta=y_p=1$. The leading contribution
 $\phi_P(0)=1$ is the unperturbed critical percolation probability and 
 the perturbation enters this expression only 
 through the sub-leading terms. 

\item Marginal perturbation, $\omega=\omega^\star=1$:\footnote{See appendix~\ref{App2} where $L$ is replaced by the distance $n$ to the first site. Although the perturbation amplitude $A$ scales in the same way, a quite different behaviour is obtained when the perturbation is marginal.}

 The scaling variable $u=A$ so that:
 \be 
 P_c(A,L)=\re^{A}\big[1+\Or\big(L^{-1}\big)\big],\quad A<0.
 \label{Pcmar}
 \ee
 The leading contribution to the critical percolation probability is  
 the universal scaling function~\eref{phiPu} of the 
 perturbation amplitude
 in agreement with~\eref{phimar} considering~\eref{xP}. 

\item Relevant perturbation, $\omega<1$:
 
 Then, $|u|$ is large but $|u/L|$ remains small. The critical percolation probability following
 from~\eref{PLu} reads:
 \be 
 P_c(A,L)=\exp\left(AL^{1-\omega}-\frac{A^2}{2}
 L^{1-2\omega}+\ldots\right).
 \label{Pcrel-1}
 \ee
 When $\omega>1/2$, the second term in the series is small 
 and leads to a correction to 
 scaling.\footnote{Otherwise a non-scaling 
 exponential factor enters the leading contribution
 to~$P_c(A,L)$.} Then, one has
 \be
 P_c(A,L)=\exp\big(AL^{1-\omega}\big) \left[1+\Or\big(L^{-2\omega+1}\big)\right],\quad 
 1/2<\omega<1,\quad A<0,
 \label{Pcrel-2}
 \ee
 in agreement with~\eref{phi} given~\eref{phiPu} since $x_\vf=0$ 
 and~$y_\Delta=1$. The critical percolation probability vanishes
 with an essential singularity as $L\to\infty$. This behaviour
 is expected for a system which is in the 
 non-percolating phase for $A<0$.
 
 The convergence of finite-size data to 
$\phi_P$ is shown in figure~\ref{fig-1}.

\end{itemize}

\section{The 1d Ising chain in a transverse field}
\label{sec:4}

In this section we consider a 1d Ising model in a transverse field~\cite{pfeuty70,kogut79} with a size-dependent perturbation of the first-neighbour interaction, looking for its influence on the surface spin magnetization. 

\subsection{Hamiltonian and surface magnetization}
The quantum chain Hamiltonian reads
\be
{\cal H}_L=-\Lambda\sum_{n=1}^{L-1}\sigma_{n}^x\sigma_{n+1}^x-\sum_{n=1}^L\sigma_n^z,
\quad \Lambda=\lambda+\frac{A}{L^\omega},
\label{ITF}
\ee
where $\sigma^x$ and $\sigma^z$ are Pauli spin operators and 
the two-spin interaction $\lambda>0$ is perturbed by the 
size-dependent term, $A/L^\omega$, favouring order (disorder) 
when $A$ is positive (negative). 

We study the scaling behaviour of the surface magnetization which is
given by the off-diagonal matrix element of the surface spin operator
\be
m_s=|\langle0|\sigma_1^x|1\rangle|,
\label{dms}
\ee
between the ground state and the first-excited state of the quantum chain, with fixed boundary condition at $L$. Using fermionic techniques it can be shown~\cite{peschel84,igloi93} that, for a finite system with size $L$, the surface magnetization 
takes the following form:
\be
m_{sL}(\Lambda)=\left(\sum_{k=0}^{L-1}\Lambda^{-2k}\right)^{-1/2}=
\left[\frac{1-(\lambda+A L^{-\omega})^{-2}}{1-(\lambda+A 
L^{-\omega})^{-2L}}\right]^{1/2}.
\label{msL}
\ee

\subsection{Unperturbed system}
In the infinite unperturbed system, i.e., when $L\to\infty$ and $A=0$, the series 
in~\eref{msL} diverges and $m_s=0$ when $\lambda\leqslant1$.
In the ordered phase, corresponding to $\lambda>1$, the surface magnetization behaves as:
\be
m_{s\infty}(\lambda)=\big(1-\lambda^{-2}\big)^{1/2}.
\label{ms}
\ee
It vanishes at the critical coupling, $\lambda_c=1$, with a surface critical exponent 
$\beta_s=1/2$. In a finite unperturbed critical system, with
$A=0$ and $\lambda=1$, \eref{msL} yields the following scaling behaviour:
\be
m_{sL}(\lambda_c)=\left(\sum_{k=0}^{L-1}1\right)^{-1/2}=L^{-1/2}.
\label{msL1}
\ee
Thus, the scaling dimension of the surface magnetization is $x_{ms}=1/2$. 
Given $\beta_s=\nu x_{ms}$, the correlation length exponent $\nu=1$ and 
the bulk thermal exponent $y_t=1/\nu=1$, too. A deviation from the 
critical coupling transforms as
\be  
\lambda'-\lambda_c=b^{y_t}(\lambda-\lambda_c)=b(\lambda-\lambda_c),
\label{lambda'}
\ee
so that, in a finite-size off-critical system, the appropriate scaling variable is:
\be 
u=L(\lambda-\lambda_c)=L(\lambda-1),\quad \lambda=1+\frac{u}{L}.
\label{ulambda}
\ee
The surface magnetization~\eref{msL} with $A=0$ involves 
\be 
\lambda^{-2}=\left(1+\frac{u}{L}\right)^{-2}
=1-\frac{2u}{L}+\frac{3u^2}{L^2}+\ldots
\label{lamb2}
\ee
and 
\be 
\lambda^{-2L}=\left(1+\frac{u}{L}\right)^{-2L}  
=\exp\left(-2u+\frac{u^2}{L}+\ldots\right)  
=\re^{-2u}\left(1+\frac{u^2}{L}+\ldots\right), 
\label{lamb2L}
\ee  
when $u^2/L\ll1$. Finally, one obtains the following 
finite-size scaling behaviour for the off-critical 
unperturbed system:  
\be
m_{sL}\left(1+\frac{u}{L}\right)=L^{-1/2}
\left(\frac{2u}{1-\re^{-2u}}\right)^{1/2}
\left[1+\left(\frac{u}{\re^{2u}-1}-\frac{3}{2}\right)\frac{u}{2L}+\ldots\right],\quad\frac{u^2}{L}\ll1.
\label{msLulamb}
\ee
The leading contribution gives the scaling function of 
the surface magnetization
\be 
\phi_{ms}(u)=\left(\frac{2u}{1-\re^{-2u}}\right)^{1/2},\quad\frac{u^2}{L}\ll1.
\label{phimsu}
\ee
The critical finite-size result in~\eref{msL1}, 
$\phi_{ms}(0)=1$, is recovered when $u\to0$.

\subsection{Perturbed system}
Since $y_t=1$, according to~\eref{Ay}, the amplitude of the thermal size-dependent perturbation  transforms as: 
\be
A'=b^{y_t-\omega}A=b^{1-\omega}A.
\label{a'}
\ee 
As above for the percolation problem, the perturbation is marginal at $\omega^\ast=1$, irrelevant when $\omega>1$ 
and relevant when $\omega<1$. 

Let us study its influence on the finite-size scaling behaviour of 
the surface magnetization for the critical value of the unperturbed coupling, $\lambda=\lambda_c$. Then, the first-neighbour interaction in~\eref{ITF} 
is~$\Lambda_c=\lambda_c+A/L^\omega$ so that, according to~\eref{ulambda} and in agreement with~\eref{a'}, the associated scaling variable is now
\be 
u=L(\Lambda_c-\lambda_c)=L^{1-\omega}A.
\label{uAms}
\ee
The critical surface magnetization
\be
m_{sc}(A,L)=m_{sL}\left(1+\frac{u}{L}\right)
=\left[\frac{1-(1+u/L)^{-2}}{1-(1+u/L)^{-2L}}\right]^{1/2},
\label{mscL}
\ee
is given by~\eref{msLulamb} and behaves in the following way in the three different regimes:

\begin{itemize}
 
 \item Irrelevant perturbation, $\omega>1$:
 
 Then, both $u$ and $u^2/L\ll1$ so that \eref{msLulamb} yields:
 \be
 \msc(A,L)= L^{-1/2}\left(1+\frac{A}{2L^{\omega-1}}+\ldots\right).
 \label{mscLirr}
 \ee
 The leading term gives the unperturbed finite-size result in~\eref{msL1}. 
 The perturbation introduces non-vanishing corrections 
 to this leading behaviour. 
 
 \item Marginal perturbation, $\omega=\omega^\ast=1$:
 
 The scaling variable is then $u=A$ and $u^2/L$ remains 
 small, thus~\eref{msLulamb} gives:
 \be
 \msc(A,L)=
 \left(\frac{2A}{1-\re^{-2A}}\right)^{1/2}L^{-1/2}\left[1+
 \left(\frac{A}{\re^{2A}-1}-\frac{3}{2}\right)\frac{A}{2L}+\ldots\right].
 \label{mscLmar}
 \ee
 The leading contribution to the surface magnetization has the same   
 finite-size scaling with $L$ as in the unperturbed system, 
 although with the varying universal amplitude $\phi_{ms}(A)$ 
 in~\eref{phimsu}. 
  
 \item Relevant perturbation, $\omega<1$:
 
 The scaling variable $u\gg1$ and the correction term 
 $u^2/L=A^2/L^{2\omega-1}$ in~\eref{lamb2L} remains small 
 only when $\omega>1/2$.
 
 When $u>0$, this non-scaling correction term is not 
 dangerous because it enters a negative exponential  
 which  can be neglected. Then, for $0<\omega<1$, the 
 critical surface magnetization following 
 from~\eref{mscL} behaves as:
 \be
 \msc(A,L)=\sqrt{2A}L^{-\omega/2}\left(1-\frac{3A}{4L^\omega}
 +\ldots\right),\quad A>0.
 \label{mscLrel-1}
 \ee
 It vanishes as $L^{-\omega/2}$
 in agreement with~\eref{phirel-3} with $x_\vf=x_{ms}=1/2$ and $y_\Delta=y_t=1$.   
 At $\omega=0$, the infinite system is in its ordered phase, with $A$ 
 giving a constant deviation from the critical coupling. 
 Thus, the surface magnetization behaves as $\msc(A,\infty)=\sqrt{2A}$ 
 and vanishes with a critical exponent $\beta_s=1/2$, as expected. 
 
 When $u<0$,~\eref{lamb2L} gives the dominant contribution 
 to the  denominator in~\eref{mscL}. When $\omega\leqslant 1/2$, 
 the effect of $u^2/L$ is to generate a non-scaling exponential factor
 which is no longer negligible. The surface magnetization keeps 
 its scaling form~\eref{mscL} only when $\omega$ belongs to the interval 
 $1/2<\omega<1$ for which
 \be
 \msc(A,L)=\sqrt{2|A|}L^{-\omega/2}
 \exp \left(-|A|L^{1-\omega}\right)\left(1-\frac{A^2}{2L^{2\omega-1}}+\ldots\right)
 ,\quad A<0.
 \label{mscLrel-2}
 \ee
 The surface magnetization vanishes with an essential 
 singularity. Such a behaviour is not surprising since 
 $A<0$ drives the system into its disordered 
 phase. Note that~\eref{mscLrel-2} has the 
 expected scaling form~\eref{phi} with $x_\vf=1/2$ and $y_\Delta=1$.

\end{itemize}

\begin{figure}[h]
	\begin{center}
		\includegraphics[width=7.5cm,angle=0]{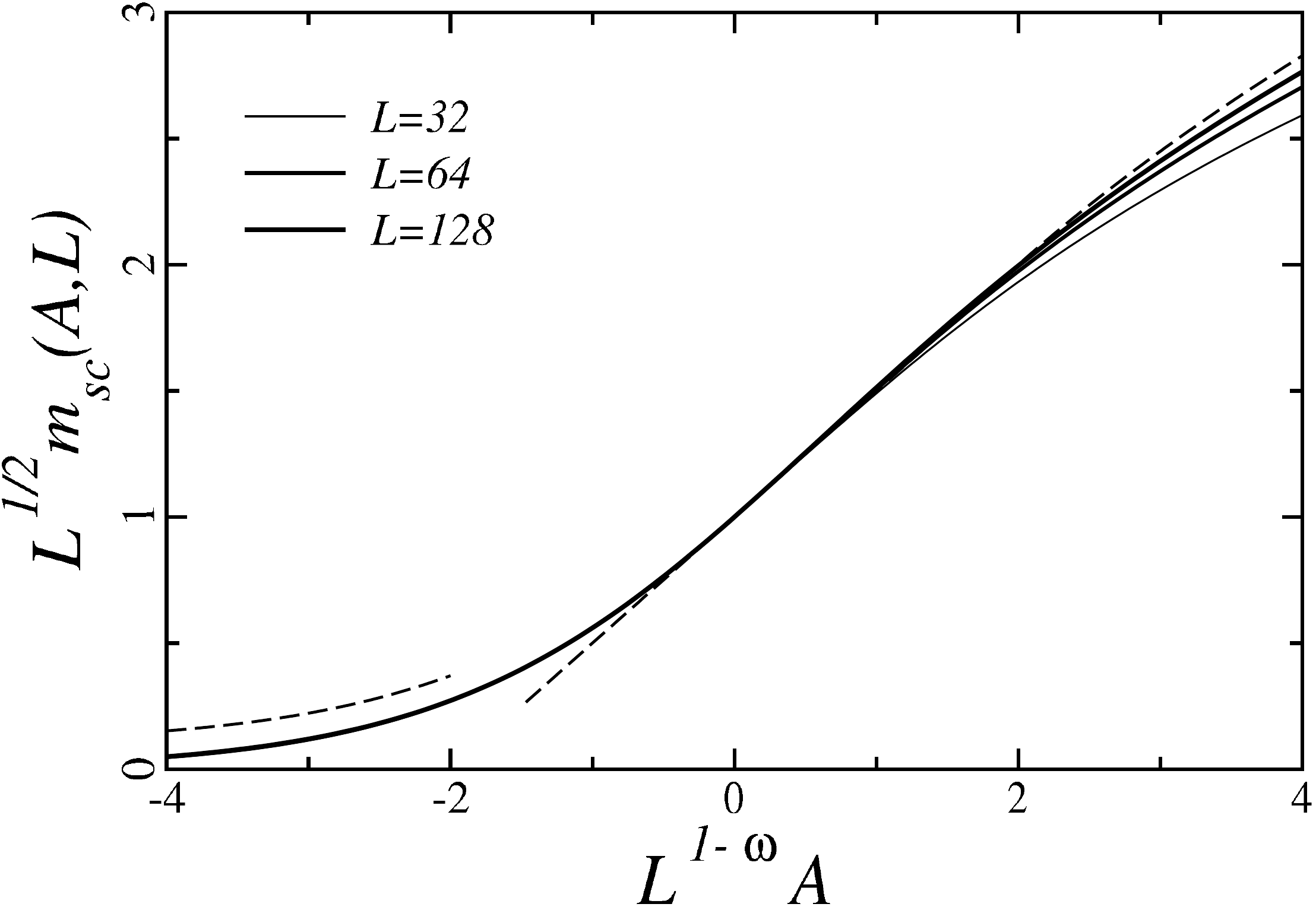}
	\end{center}
	\vglue -.5cm
	\caption{Evolution of the scaled critical surface magnetization in~\eref{mscL} towards the universal scaling 
		function $\phi_{ms}(u)$ in~\eref{phimsu}, for increasing values of the size $L$. The dashed lines indicate the expected behaviours for small and large values of the scaled amplitude, $u=L^{1-\omega}A$. For $A\ll0$, it has been shifted upwards to be visible.}
	\label{fig-2} 
\end{figure}

The convergence of finite-size data to 
$\phi_{ms}$ is shown in figure~\ref{fig-2}.

 %%%%%%%%%% FIG 2  %%%%%%%%%%%%%%%%%%%%%%%%%%%%%%%
%%%%%%%%%%%%%%%%%%%%%%%%%%%%%%%%%%%%%%%%%

\section{Ising model on the fully-connected lattice}
\label{sec:5}

Let us finally consider another exactly solvable example provided by the Ising model on the fully-connected lattice with $N$ sites. 

\subsection{Hamiltonian, Ginzburg-Landau expansion and order parameter}

The perturbed Hamiltonian takes the following form
\be
-\beta{\cal H}=\CK\left(\frac{1}{2N}\sum_{i\neq j}\sigma_i\sigma_j\right),
\quad \CK=K+\frac{C}{N^\omega},
\quad K=\beta J=\frac{J}{\kBT},
\label{Hfc}
\ee
where $J>0$ is the ferromagnetic interaction between the Ising spins $\sigma_i=\pm1$. The 
factor $1/N$ is needed to have a total energy proportional to $N$ in the thermodynamic limit.
The two-spin interaction is perturbed by the size-dependent term $C/N^\omega$.

Introducing the total magnetization $M=\sum_{i=1}^N\sigma_i$ so that 
$\sum_{i\neq j}\sigma_i\sigma_j=M^2-N$, the partition function is given by:
\be
Z_N=\Tr_{\{\sigma\}}\re^{-\beta{\cal H}}=\re^{-\CK/2}
\Tr_{\{\sigma\}}\exp\left(\frac{\CK M^2}{2N}
\right).
\label{ZN-1}
\ee
Making use of the Stratonovich-Hubbard 
transformation~\cite{stratonovich57,hubbard59}
\be
\exp\left(\frac{\CK M^2}{2N}\right)=\left(\frac{N}{2\piup}\right)^{1/2}
\int_{-\infty}^{+\infty}\rd\eta\, \exp\left(-\frac{N}{2}\eta^2+\CK^{1/2}\eta M\right),
\label{id}
\ee
the partition function can be rewritten as:
\be
Z_N=\left(\frac{N}{2\piup}\right)^{1/2}\!\! \re^{-\CK/2}\!\!
\int_{-\infty}^{+\infty}\!\!\! \rd\eta\, \exp\left(-\frac{N}{2}\eta^2\right)
\prod_{i=1}^N\Tr_{\sigma_i}\re^{\CK^{1/2}\eta\sigma_i}.
\label{ZN-2}
\ee
Summing on the spin states leads to
\be
Z_N=2^N\left(\frac{N}{2\piup}\right)^{1/2}
\re^{-\CK/2}\int_{-\infty}^{+\infty}\!\!\! \rd\eta\ \re^{-Nf(\CK,\eta)},
\label{ZN-3}
\ee
where 
\be
f(\CK,\eta)=\frac{\eta^2}{2}-\ln\cosh\left(\CK^{1/2}\eta\right)
\label{feta-1}
\ee
is the free energy per site.

Close to the critical point, the free energy density in~\eref{feta-1} 
can be expanded in even powers of $\eta$ as:
\be
f(\CK,\eta)=-\left(K+\frac{C}{N^\omega}-1\right)\frac{\eta^2}{2}
+\left(K+\frac{C}{N^\omega}\right)^2\frac{\eta^4}{12}+\Or(\eta^6).
\label{feta-2}
\ee
Taking $|\eta|$ for the order parameter, its mean value is given by:
\be
m_N(\CK)=\frac{\int_0^\infty\eta\,\re^{-Nf(\CK,\eta)}\rd\eta}{\int_0^\infty
\re^{-Nf(\CK,\eta)}\rd\eta}.
\label{mN}
\ee

\subsection{Unperturbed system}

In the unperturbed system $C=0$, and~\eref{feta-2} reduces to:
\be
f(K,\eta)=-(K-1)\frac{\eta^2}{2}+K^2\,\frac{\eta^4}{12}+\Or(\eta^6).
\label{fK}
\ee
In the thermodynamic limit, as $N\to\infty$, the order parameter 
in~\eref{mN} is given by the value of $\eta$ minimizing the free energy. It
is non-vanishing when $K>K_c=1$ where
\be 
m_\infty(K)\approx\sqrt{3(K-1)},\quad K>1.
\label{minfK}
\ee
As expected for a fully-connected lattice where a spin 
interacts with all the others, the Ising mean-field critical behaviour, 
$\beta=1/2$, is obtained.

At the critical point, the free energy density is given by:
\be
f(K_c,\eta)=\frac{\eta^4}{12}+\Or \big(\eta^6 \big).
\label{fKc}
\ee
The change of variable $t=N\eta^4/12$ in~\eref{mN} 
immediately leads to:
\be 
m_N(K_c)=\left(\frac{12}{N}\right)^{1/4}\frac{\int_0^\infty 
t^{-1/2}\re^{-t}\rd t}{\int_0^\infty t^{-3/4}\re^{-t}\rd t}
=\frac{12^{1/4}\sqrt{\piup}}{\Gamma(1/4)}N^{-1/4}.
\label{mNKc}
\ee
This finite-size scaling with $N$ is an old result, first obtained by Kittel~\cite{kittel65}.
It was later remarked~\cite{botet82,botet83} that~\eref{mNKc} is 
actually the standard scaling behaviour for the mean-field Ising model if one relates the number $N$ of sites in the fully-connected lattice to an effective size $L$ through $N=L^{d_c}$. In this relation, $d_c$ is the upper critical dimension above which the mean-field behaviour is obtained with short-range interactions.
For the Ising model, $d_c=4$, so that:
\be 
L=N^{1/d_c}=N^{1/4}.
\label{L}
\ee
Using this relation in~\eref{mNKc} and $\beta=x_m/y_t=1/2$ yields the scaling dimensions of the mean-field Ising model:
\be 
x_m=1,\quad y_t=1/\nu=2.
\label{scaldim}
\ee

Let us now consider a finite-size off-critical system. According to~\eref{L}
and~\eref{scaldim}, the scaling variable takes the following form:
\be 
v=L^{y_t}(K-K_c)=N^{y_t/d_c}(K-K_c)=N^{1/2}(K-1),
\quad K=1+\frac{v}{N^{1/2}}.
\label{v-1}
\ee
The scaling of the order parameter in~\eref{mNKc} suggests the change of variable
$\eta=xN^{-1/4}$ in~\eref{fK}. As a function of $x$ and $v$, the free energy density takes the following form:
\be 
Nf(1+v/N^{1/2},xN^{-1/4})=Ng_N(v,x)
=-v\frac{x^2}{2}+\frac{x^4}{12}
+\frac{vx^4}{6N^{1/2}}+\Or \big(N^{-1}\big).
\label{gvx}
\ee
In this expansion, the non-scaling third term and the 
following terms can be 
neglected, so that~\eref{mN} yields:
\be
m_N \left(1+v/N^{1/2}\right)=N^{-1/4}\frac{\int_0^\infty x\,\re^{-Ng_N(v,x)}\rd x}{\int_0^\infty
 \re^{-Ng_N(v,x)}\rd x}
 = N^{-1/4}\frac{J_1(v)}{J_0(v)}\left[1+\Or\!\left(vN^{-1/2}\right)\right].
 \label{mvN}
 \ee
$J_n(v)$ is given by
\be
J_n(v)=\int_0^\infty\! x^n\exp\left(-\frac{x^4}{12}+\frac{vx^2}{2}\right)\rd x
=\frac{1}{2}\,6^{(n+1)/4}\Gamma\left(\frac{n+1}{2}\right)
\re^{3v^2/8}D_{-(n+1)/2}\left(-\sqrt{\frac{3}{2}}v\right),
\label{Jn}
\ee
where
\be
D_\mu(z)=\frac{\exp\left(-{z^2}/{4}\right)}{\Gamma(-\mu)}
\int_0^\infty \re^{-zx-x^2/2}x^{-\mu-1}\rd x,\quad [\Re(\mu)<0],
\label{Dmuz}
\ee
is a parabolic cylinder function \cite{gradshteyn80}. Thus, \eref{mvN} 
can be rewritten as:
\be
m_N(1+v/N^{1/2})=\frac{1}{\sqrt{\piup}}
\frac{D_{-1}\left(\!-\sqrt{\frac{3}{2}}v\!\right)}
{D_{-1/2}\left(\!-\sqrt{\frac{3}{2}}v\!\right)}
\left(\!\frac{6}{N}\!\right)^{1/4}\left[1+\Or\!\left(N^{-1/2}\right)\right].
\label{mvN-1}
\ee
The off-critical finite-size behaviour of the order parameter 
has the standard scaling form $N^{-1/4}\phi_m(v)$ with 
a scaling function given by:
\be
\phi_m(v)=\frac{6^{1/4}}{\sqrt{\piup}}
\frac{D_{-1}\left(-\sqrt{\frac{3}{2}}v\right)}
{D_{-1/2}\left(-\sqrt{\frac{3}{2}}v\right)}
,\quad v=N^{1/2}(K-1).
\label{phim}
\ee
The different limiting behaviours of the ratio $R(v)$ of parabolic cylinder functions entering~\eref{phim} are studied in appendix~\ref{paracyl}.

\subsection{Perturbed system}

Let us now consider a perturbed critical system with $K=K_c=1$ and $C=\Or(1)$. 
The size-dependant perturbation in~\eref{Hfc} can be rewritten as $C/L^{d_c\omega}$ and,
according to~\eref{Ay} and~\eref{scaldim}, its amplitude transforms as:
\be 
C'=b^{y_t-d_c\omega}C=b^{2-4\omega}C.
\label{C'}
\ee
The finite-size scaling variable is then
\be
v=L^{2-4\omega}C=N^{\frac{1}{2}-\omega}C,
\label{v_2}
\ee
leading to the following scaling behaviour for the critical 
magnetization:
\be
m_c(C,N)=m_N \left(1+C/N^\omega \right) =m_N \left(1+v/N^{1/2}\right)=N^{-1/4}\phi_m(v),
\quad v=N^{\frac{1}{2}-\omega}C.
\label{mcN-0}
\ee
According to~\eref{C'}, the perturbation is irrelevant above $\omega=\omega^\ast=1/2$, marginal at $\omega^\ast$,  
and relevant below. Let us now look at the finite-size behaviour of 
the order parameter in the perturbed critical system in these three regimes:

\begin{itemize}
\item Irrelevant perturbation, $\omega>1/2$:
 
 With $C=\Or(1)$ and $N\gg1$, the scaling variable $v$ is small 
 and the non-scaling correction term in~\eref{gvx}, behaving as $Cx^4/N^\omega$,
 can be neglected.
 A first-order expansion~\eref{R-3} of the ratio of parabolic cylinder 
 functions $R(v)$ in the scaling function~\eref{phim} yields: 
  \be
 m_c(C,N)\approx\frac{\sqrt{\piup}}{\Gamma(1/4)}\left(\frac{12}{N}\right)^{1/4}
 \left\{1+\left[\frac{1}{\sqrt{\piup}}-\frac{\Gamma(3/4)}{\Gamma(1/4)}\right]
 \frac{\sqrt{3}C}{N^{\omega-1/2}}\right\}.
 \label{mcN-1}
 \ee
 The perturbation only contributes a correction to the $N^{-1/4}$ scaling behaviour 
 of the unperturbed system.

  \item Marginal perturbation, $\omega=\omega^\ast=1/2$: 

 Then, $v=C$, so that the first correction term in~\eref{gvx} remains negligible.
 Thus, the leading contribution to $m_c$ reads:
 \be 
 m_c(C,N)\approx\phi_m(C)N^{-1/4}.
 \label{mcN-2}
 \ee
 One recovers the unperturbed scaling with $N$ with a $C$-dependent
 universal amplitude.

 \item Relevant perturbation, $\omega<1/2$:
 
 With a relevant perturbation, $|v|$ is large and the leading 
 contribution to $R(v)$ depends 
 on the sign of the scaling variable. Using~\eref{R-4} and~\eref{R-5},
 one obtains:
 \be
 m_c(C,N)\approx
 \left\{
 \begin{array}{ll}
 \sqrt{3C}N^{-\omega/2},&\quad C>0,\\
 \sqrt{\frac{2}{\piup|C|}}N^{-(1-\omega)/2},&\quad C<0.
 \end{array}
 \right.
 \label{mcN-3}
 \ee
 As before in the relevant case, the scaling exponent depends on 
 $\omega$. This new scaling behaviour may lead to a dangerous 
 non-scaling correction to the free energy density in~\eref{gvx}
 since $x$ is no longer dimensionless. The problem is studied 
 in appendix~\ref{validity} where it is shown that~\eref{mcN-3} 
 remains valid on the interval~$1/3<\omega<1/2$ when $C>0$ 
 and without restriction when $C<0$ and $\omega<1/2$.

\end{itemize}

\begin{figure}[h]
 	\begin{center}
 		\includegraphics[width=8cm,angle=0]{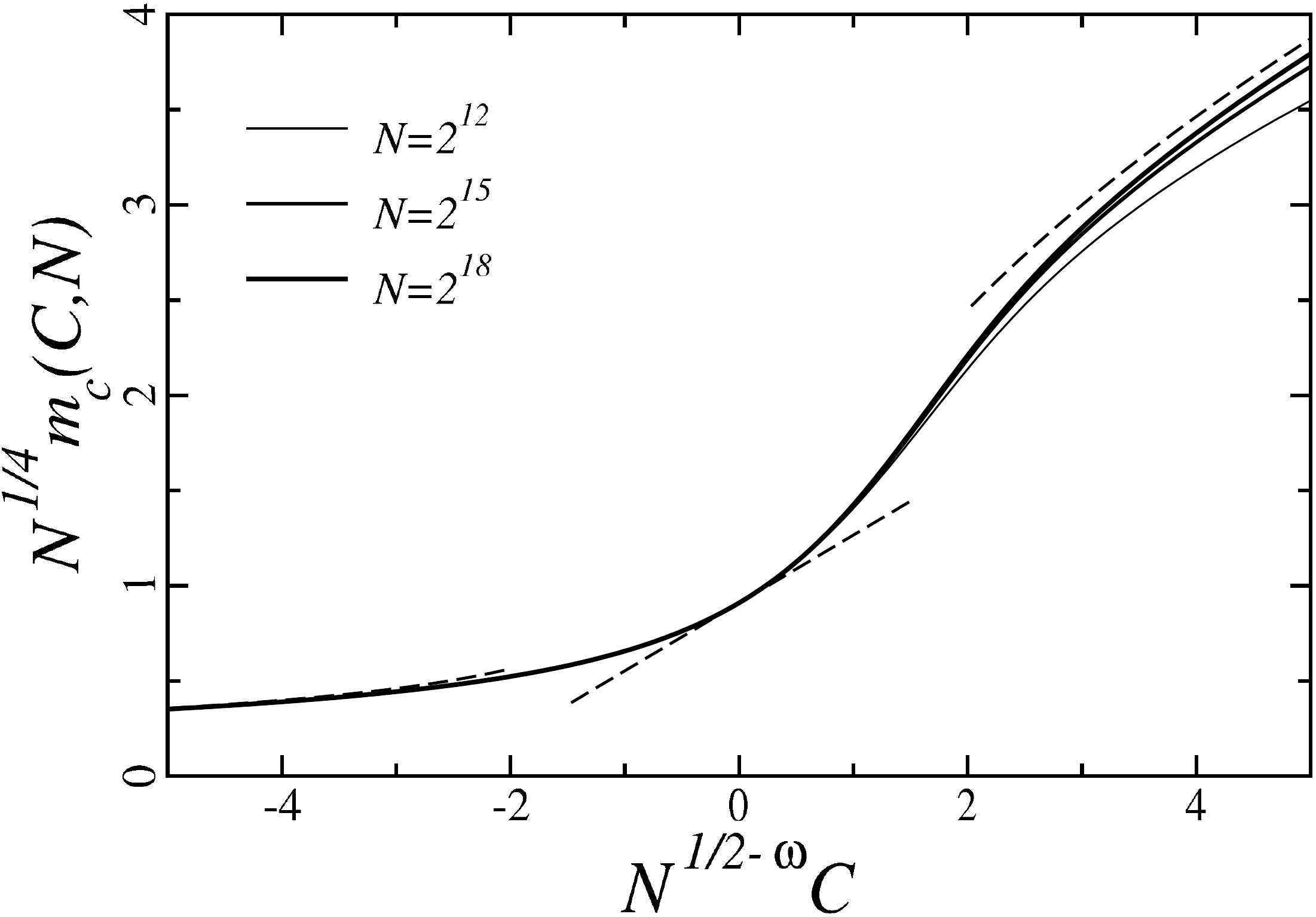}
 	\end{center}
 	\vglue -.5cm
 	\caption{Evolution of the scaled critical bulk magnetization, given by~\eref{feta-2} and~\eref{mN} with $K=1$, towards the universal scaling 
 		function $\phi_m(v)$ in~\eref{phim}, for increasing values of the number of spins $N$. The dashed lines indicate the expected behaviours for small and large values of the scaled amplitude, $v=N^{({1}/{2})-\omega}C$. The convergence is quite slow for large positive values of $v$.}
 	\label{fig-3}
 \end{figure}

The convergence of finite-size data to 
 $\phi_m$ is shown in figure~\ref{fig-3}.

 %%%%%%%%%% FIG 3  %%%%%%%%%%%%%%%%%%%%%%%%%%%%%%%
 %%%%%%%%%%%%%%%%%%%%%%%%%%%%%%%%%%%%%%%%%

\section{Discussion}
\label{sec:6}

Let us first consider the finite-size scaling behaviour which is obtained in~\eref{phimar} for a truly marginal perturbation. Instead of a scaling dimension continuously varying with $A$, the behaviour is the same as in the unperturbed system in~\eref{phiun}. 

A continuously varying critical exponent would be associated with a line of fixed points in the critical surface. By definition, the critical surface is the set of points in the parameter space where the correlation length diverges. Its mere existence requires an infinite-size
system for which the size-dependent 
perturbation~\eref{Delta} vanishes. Thus, the critical Hamiltonians and its associated fixed points remain the unperturbed ones, which explains the observed scaling behaviour.                                 

The marginalism affects the amplitude which is continuously varying with $A$. Since 
$\phi_\vf(A)=\Phi_c(A,1)$, where $\Phi_c$ is a finite-size scaling function, the variation of the amplitude with $A$ is universal.

For a marginally perturbed finite-size 2d system in the cylinder geometry, the lowest gap, $G_\vf(\Delta,L)$, which is an inverse correlation length associated with the local operator $\vf$, either the magnetization or the energy density, scales as:
\be
G_\vf(\Delta',L')=G_\vf(b^{y_\Delta}\Delta,L/b)=bG_\vf(\Delta,L),
\quad \Delta=\frac{A}{L^{y_\Delta}}\,.
\label{Gsca}
\ee
With $b=L$, one obtains:
\be
G_\vf(\Delta,L)=G_\vf(A,1)L^{-1}.
\label{Gsca1}
\ee
Thus, comparing to \eref{gapexpo} in appendix~\ref{conf}, 
the universal amplitude of the gap is given by
\be
G_\vf(A,1)=2\piup x_\vf(A),
\label{Gsca2}
\ee
where $x_\vf(A)$ is the varying local exponent 
associated with the marginal radial defect from which the 
homogeneous size-dependent perturbation $\Delta$ is the conformal transform on the cylinder~\cite{bariev91,bariev91a,turban91}.

\appendix

\section{Marginal size-dependent perturbation resulting from a conformal transformation\label{conf}}
In this section we show how a homogeneous marginal size-dependent perturbation may result from the conformal transformation of a radial marginal perturbation in a 2d
infinite system~\cite{bariev91,bariev91a,turban91,igloi93}.

The original system is critical and the radial marginal perturbation 
\be
\Delta=\frac{A}{(2\piup r)^{y_\Delta}}
\label{Delta-r}
\ee
acts on the local operator $\zeta$, in the plane with polar coordinates $(r,\theta)$, so that $y_\Delta=2-x_\zeta$. 

Under the conformal transformation~\cite{cardy84}
\be
w(z)=\frac{L}{2\piup}\ln z,\quad 
z=r\re^{\ri\theta},\quad
w=x+\ri y,
\label{ct}
\ee
the $z$-plane is mapped onto a $w$-cylinder with
\be
x=\frac{L}{2\piup}\ln r,\quad 
-\infty<x<+\infty,\quad
y=\frac{L\theta}{2\piup},\quad 0\leqslant y<L.
\label{xy}
\ee
The associated local dilation factor is
\be 
b(z)=|w'(z)|^{-1}=2\piup\frac{r}{L},
\label{bz}
\ee
so that the radial perturbation transforms into
\be
\Delta'=[b(z)]^{y_\Delta}\Delta=\frac{A}{L^{y_\Delta}},
\label{Delta-w}
\ee
i.e., a constant size-dependent deviation from the critical point.

Let  $\vf$ be an operator which, in the original system, displays a varying critical exponent $x_\vf(A)$ under the influence of the radial marginal perturbation.
Comparing the expression of the critical two-point correlation function $\langle\vf(\rho_1,\theta_1)\vf(\rho_2,\theta_2)\rangle$ in the original system to the form obtained by transforming back  $\langle\vf(x_1,y_1)\vf(x_2,y_2)\rangle$ on the cylinder, one can show that $x_\vf(A)$ is given, up to a constant factor, by the $A$-dependent universal amplitude of the associated lowest gap, $G_\vf=E_\vf-E_0$, in the transformed system with a size-dependent marginal perturbation~\cite{bariev91,bariev91a,turban91,igloi93}:
\be
x_\vf(A)=\frac{L}{2\piup}G_\vf.
\label{gapexpo}
\ee

\section{Hilhorst--van Leeuwen perturbation of percolation in 1d}
\label{App2}

We consider the 1d bond percolation model on a finite 
system with size $L$ with a perturbation of the Hilhorst-van Leeuwen type~\cite{hilhorst81}.
The bond occupation probability is a decreasing function 
of the distance~$n$ to the left-hand surface:
\be 
\pi_n=p+\Delta_n=p+\frac{A}{n^\omega},\quad\omega>0
,\quad -p\leqslant A\leqslant 1-p.
\label{pin}
\ee
Evidently, the scaling behaviour of the perturbation amplitude~$A$ is the same as in~\eref{scalA} but the perturbation now involves the set of $\Delta_n$ $(n=1,L)$ instead of a single one, $\Delta_L$. We shall only consider the marginal case with $\omega=1$. The critical percolation probability is then given by:
\be 
P_c(A,L)=\prod_{n=1}^L\left(1+\frac{A}{n}\right)
=\frac{1}{L!}\frac{\Gamma(L-|A|+1)}{\Gamma(1-|A|)},
\quad -1\leqslant A \leqslant 0.
\label{PcHvL-1}
\ee
When $L$ is large, Stirling's formula~\cite{abramowitz72} yields:
\be 
\ln\left[\frac{\Gamma(L-|A|+1)}{\Gamma(L+1)}\right]=-|A|\ln L
+\Or\big(L^{-1}\big).
\label{lnGamma}
\ee
Finally, the critical percolation probability displays a traditional marginal behaviour
\be 
P_c(A,L)\approx\frac{L^{-|A|}}{\Gamma(1-|A|)},
\label{PcHvL-2}
\ee
with a perturbation-dependent finite-size scaling exponent, 
$x_P(A)=|A|$.

\section{Limiting behaviour of the ratio of parabolic cylinder functions}
\label{paracyl}
Let us study the asymptotic behaviour for small and large values of $|v|$ of
\be 
R(v)=\frac{D_{-1}\left(\!-\sqrt{\frac{3}{2}}v\!\right)}
 {D_{-1/2}\left(\!-\sqrt{\frac{3}{2}}v\!\right)}
\label{R-1}
\ee
in~\eref{phim}.
\begin{itemize}
 \item $|v|\ll1$:
 
 The relation $D_{-r-1/2}(x)=U(r,x)$ and the known values of $U(r,0)$ 
 and $U'(r,0)$~\cite{abramowitz72} lead to the expansion
 \be
 D_\mu(x)=\frac{2^{\mu/2}\sqrt{\piup}}{\Gamma\!\left(\frac{1-\mu}{2}\right)}
 \left[1-\frac{\sqrt{2}\,\Gamma\!\left(\frac{1-\mu}{2}\right)}
 {\Gamma\!\left(-{\mu}/{2}\right)}x+\Or\big(x^2\big)\right],
 \label{Dmux-1}
 \ee
 so that:
 \be
 R(v)=\frac{\Gamma(3/4)}{2^{1/4}}
 \left[1+\left(\frac{1}{\sqrt{\piup}}-\frac{\Gamma(3/4)}{\Gamma(1/4)}\right)
 \sqrt{3}v+\Or\big(v^2\big)\right].
 \label{R-2}
 \ee
 Making use of the reflection formula, $\Gamma(3/4)\Gamma(1/4)=\sqrt{2}\,\piup$,
 one obtains
 \be
 R(v)=\frac{2^{1/4}\piup}{\Gamma(1/4)}
 \left[1+\left(\frac{1}{\sqrt{\piup}}-\frac{\Gamma(3/4)}{\Gamma(1/4)}\right)
 \sqrt{3}v+\Or\big(v^2\big)\right].
 \label{R-3}
 \ee

 \item $v\gg1$:
 
 For $x\gg1$, the parabolic cylinder function behaves as~\cite{gradshteyn80}
 \be
 D_\mu(-x)\sim \frac{\sqrt{2\piup}}{\Gamma(-\mu)}\re^{x^2/4}x^{-\mu-1}
 \left[1+\Or\big(x^{-2}\big)\right],
 \label{Dmux-2}
 \ee
 leading to:
 \be
 R(v)=\left(\frac{3}{2}\right)^{1/4}\sqrt{\piup v}\left[1+\Or\big(v^{-2}\big)\right].
 \label{R-4}
 \ee

 \item $-v\gg1$:
 
 For $x\gg1$, one has~\cite{gradshteyn80}
 \be
 D_\mu(x)\sim\,\re^{-x^2/4}x^{\mu}
 \left[1+\Or\big(x^{-2}\big)\right],
 \label{Dmux-3}
 \ee
 so that:
 \be
 R(v)=\left(\frac{2}{3}\right)^{1/4}|v|^{-1/2}\left[1+\Or\big(|v|^{-2}\big)\right].
 \label{R-5}
 \ee
 
\end{itemize}

\section{Limits on the validity of \eref{mcN-3}}
\label{validity}

When $C>0$ the dimensionless variable $x_+$ can be defined through 
$\eta=x_+N^{-\omega/2}$. Then, according to~\eref{feta-2}, 
$Nf(1+C/N^{\omega},x_+N^{-\omega/2})$ yields the critical free 
energy of the perturbed system:
\be
Ng_N(C,x_+)
=N^{1-2\omega}\left(-\frac{Cx_+^2}{2}+\frac{x_+^4}{12}\right)
+\frac{Cx_+^4}{6N^{3\omega-1}}+\Or\big(N^{1-4\omega}\big).
\label{gCx+}
\ee
It follows that the correction term is indeed small when $\omega>1/3$. 
In the interval $1/3<\omega<1/2$ the leading contribution to the 
order parameter in~\eref{mN} is given by  
\be
m_c(C,N)=N^{-\omega/2}\frac{K_1(C,N)}{K_0(C,N)},
\label{mcN-4}
\ee
where:
\be
K_n(C,N)=\int_0^\infty x^n\exp\left[N^{1-2\omega}
\left(-\frac{x^4}{12}\!+\!\frac{Cx^2}{2}\right)\right] \rd x.
\label{Kn-1}
\ee
This integral is easily evaluated using Laplace method which gives:  
\be
K_n(C,N)\approx(3C)^{n/2}\exp \left({\frac{3}{4}N^{1-2\omega}C^2}\right)
\sqrt{\frac{\piup}{N^{1-2\omega}C}},\quad C>0.
\label{Kn-2}
\ee
As expected, \eref{mcN-4} gives back the first expression of $m_c(C,N)$ 
in~\eref{mcN-3}.

Similarly, when $C<0$ the dimensionless variable is $x_-$ such that 
$\eta=x_{-}N^{-(1-\omega)/2}$ and the critical free energy now reads:
\be
Ng_N(C,x_-)
=\frac{|C|x_-^2}{2}+\frac{x_-^4}{12N^{1-2\omega}}
-\frac{|C|x_-^4}{6N^{1-\omega}}+\Or\big(N^{-1}\big).
\label{gCx-}
\ee
Thus, with $\omega<1/2$ the first term alone survives.
With the change of variable
\be 
t=\frac{|C|x_-^2}{2}=\frac{1}{2}|C|N^{1-\omega}\eta^2
\label{txeta}
\ee
in~\eref{mN}, the second expression of $m_c(C,N)$ in~\eref{mcN-3} 
is easily recovered.

\ukrainianpart

\title{Скейлінгова поведінка під впливом однорідного збурення, що залежить від розміру}
\author{\framebox{Л. Тюрбан}}

\address{Університет Лотарингії, CNRS, F-54000 Нансі, Франція}
\makeukrtitle 

\begin{abstract}
	\tolerance=3000%
	Вивчається скінченно-вимірна скейлінгова поведінка в критичній точці, що є результатом додавання однорідного збурення, залежного від розміру, яка спадає як обернена степінь розміру системи.
	Теорія скейлінгу спочатку формулюється у рамках загального підходу, а потім ілюструється на прикладі трьох конкретних задач, для яких отримано точні результати.
	
	
\end{abstract}

\lastpage
\end{document}